# Substantial regional variation in substitution rates in the human genome: importance of GC content, gene density and telomere-specific effects

Peter F Arndt[1], Terence Hwa[2], and Dmitri A Petrov[3]

[1] Max Planck Institute for Molecular Genetics, Ihnestr. 73, 14195 Berlin, Germany
[2] Department of Physics and Center for Theoretical Biological Physics
UC San Diego, 9500 Gilman Drive, La Jolla, CA 93092-0374
[3] Department of Biological Sciences, 371 Serra St., Stanford University, Stanford, CA 94305

Corresponding author:

Peter F Arndt, email: arndt@molgen.mpg.de

## *Abstract*

This study presents the first global, 1 Mbp level analysis of patterns of nucleotide substitutions along the human lineage. The study is based on the analysis of a large amount of repetitive elements deposited into the human genome since the mammalian radiation, yielding a number of results that would have been difficult to obtain using the more conventional comparative method of analysis. This analysis revealed substantial and consistent variability of rates of substitution, with the variability ranging up to 2-fold among different regions. The rates of substitutions of C or G nucleotides with A or T nucleotides vary much more sharply than the reverse rates suggesting that much of that variation is due to differences in mutation rates rather than in the probabilities of fixation of C/G vs. A/T nucleotides across the genome. For all types of substitution we observe substantially more hotspots than coldspots, with hotspots showing substantial clustering over tens of Mbp's. Our analysis revealed that GC-content of surrounding sequences is the best predictor of the rates of substitution. The pattern of substitution appears very different near telomeres compared to the rest of the genome and cannot be explained by the genome-wide correlations of the substitution rates with GC content or exon density. The telomere pattern of substitution is consistent with natural selection or biased gene conversion acting to increase the GC-content of the sequences that are within 10-15 Mbp away from the telomere.

## *Introduction*

Mammalian genomes are highly structured entities. The earliest cytogenetic studies of the human chromosomes revealed the presence of bands of differential staining with Giemsa dye (Caspersson et al. 1971). Later studies, pioneered by G. Bernardi and colleagues, found striking variability in GC-content across the genome (Bernardi 2000; Eyre-Walker and Hurst 2001; Filipski et al. 1973). The GC-content variation is surprisingly parceled into long (100's of kb) regions of fairly similar GC-content. Moreover, genes and transposable elements are also non-randomly distributed across the genome. Most genes are found in GC-rich regions (see (Lercher et al. 2003) and references therein for a detailed analysis), while different classes of transposable elements tend to have preferences for either GC-rich or GC-poor regions. Overlaying these patterns, there is also variability in proportion of imprinted genes, methylation density, recombination rate, and chromatin structure. Understanding the evolution, functional significance of these patterns and of their interrelationships presents a major challenge of modern genomic science.

It is important to understand better the variability in the patterns of point substitution, including patterns of point mutation and biases in fixation probabilities due to natural selection or biased gene conversion. These patterns are key to gaining insight into the structure and function of our genome for a number of reasons. First of all, they may have been directly responsible for some of the observed genomic heterogeneity. For instance, the variation in these patterns is likely to have generated the observed GC-content variation, which then could have led to variability in chromatin structure and gene density distribution. We also need to understand the variability in the rates of point substitution to be able to differentiate among those genomic regions that are unusually conserved due to natural selection preserving their function and those conserved because they happen to be in regions of low rate of point substitutions. Attempts to find functional sequences by identifying conserved blocks of sequence are easily misled by even subtle variability in mutation rates (Ellegren et al. 2003).

The question of variability in the rate of point substitution remains controversial, with some authors insisting that rates of neutral substitution are approximately constant across the genome (Kumar and Subramanian 2002; Subramanian and Kumar 2003) and other arguing for 2-3 fold variation in the substitution rate (Hardison et al. 2003; Smith et al. 2002; Waterston et al. 2002). These estimates of substitution rates are based on human-mouse comparisons or do not distinguish substitutions due to different substitutional processes. Both methods have their limitations, since they either cannot distinguish changes in the human or mouse lineage, do not allow to draw a map of the relative substitution rates on the 1 Mbp scale, or wont be able to detect raised substitution rates for the G:C → X:Y or A:T → X:Y processes only. This debate is further complicated by the fact that the human genome appears to be far from equilibrium in respect to the pattern of neutral substitution. In particular it appears that the human genome is losing its GC-rich regions (Arndt et al. 2003b; Duret et al. 2002). The change in the pattern of neutral substitution apparently occurred ~90 MYA, roughly simultaneously with the radiation of eutherian mammals (Arndt et al. 2003b).

Most attempts of investigating variability in the rates of point substitution across the human genome have employed some form of genomic sequence comparison (Ellegren et al. 2003; Lercher et al. 2004). The idea is to align syntenic regions in two mammalian genomes and to infer patterns of neutral substitution from the tabulated differences between the two regions. This approach is very powerful but it does have a number of limitations. The difficulties include problems with inferring the ancestral sequence and separating changes into those that occurred in different lineages. This is especially problematic given the non-equilibrium state of the human genome and the fact that the dominant mutation process (occurring at the 50-fold higher rate than an average transversion (Arndt et al. 2003b)) in mammalian genome is a neighbor- and methylation-dependent transition of CpG to TpG/CpA. This means that all reversible and stationary models of nucleotide evolution are inappropriate in this case. Attempts to avoid these problems through exclusion of potential CpG ancestral sites or genomic positions where non-stationarity can be detected are unsatisfying for a number of reasons. Such approaches necessarily do not give us the full picture of the substitutional processes. Exclusion of CpG sites is particularly problematic, given how powerful this process is and given that even after the exclusion of such sites the CpG process is still expected to affect estimates of other rates (Arndt et al. 2003a).

Finally, in order to understand patterns of substitution unaffected by selection on gene function, it is important to limit the analysis to the best extent possible to the *functionally unconstrained* sequences such as pseudogenes or dead copies of transposable elements. One study (Hardison et al. 2003) has used sequences of human and mouse genomes and focused in part on the sequences of ancestral repeats – e.g. those copies of repetitive sequences for which we can be reasonably certain that they inserted prior to the split of human and mouse evolutionary lineages and which can be found in orthologous positions in both genomes. Unfortunately, the number of such repeats is quite limited and the analysis could be at best done at a 5 Mbp scale (Hardison et al. 2003).

In our previous study (Arndt et al. 2003b) we showed that it is possible to use sequences of the presumably neutral copies of extremely abundant Alu and Line1 transposable elements (TEs) to infer patterns of point substitution by comparing the sequences of individual TEs with the sequence of the inferred ancestral TE sequence. Moreover, by focusing on individual TE families it is possible to investigate the evolutionary history of the substitutional patterns. In this study we extend this approach to investigate the regional variation of substitution rate across the human genome. In particular, we focus on the Alu elements that have been prolific in the human lineage since mammalian radiation ~90 MYA (Britten et al. 1988; Jurka and Smith 1988; Kapitonov and Jurka 1996) and now constitute upward of 10% of the human genome. Because they are so spectacularly abundant, practically any given 1 Mbp region of the human genome contains enough Alu copies to provide us with sufficiently precise estimates of the patterns and history of point substitution in that region. (Qualitatively similar results are obtained using those L1 elements that were inserted into the human genome after the mammalian radiation. However, the amount of the mammalian L1 elements (39 Mbp) is about 8 times less than the Alu sequences (312 Mbp). Therefore the L1 analysis cannot be carried out at the same 1 Mbp resolution with the same statistical significance and only the results derived from the Alu elements are presented.) Through the analysis of Alu and L1 elements, we find evidence that rates of neutral substitution vary by 2-3 fold across the genome and investigate the determinants of this variation. We confirm that GC-content is the major predictor of the rate of point substitution. We also show that gene-poor regions have particularly high rates of transversions away from C:G pairs. Finally, we demonstrate that telomeric regions have an unusual GC-enriching pattern of substitution, qualitatively different from other patterns in the rest of the genome.

## *Material and Methods*

**Genomic data**

Our analysis is based on human genomic sequence data (NCBI 33 assembly from April 1st, 2003), which is available at www.ensembl.org (Hubbard et al. 2002). We used the RepeatMasker (http://www.repeatmasker.org) to identify copies of various repetitive elements (REs). The ancestral master sequences of the REs were taken from RepBase (Jurka 2000). By default the RepeatMasker uses different scoring systems to identify and align REs in different isochores. To exclude potential biases of our estimation of substitution frequencies due to different scoring systems the RepeatMasker was forced to use only the scoring system optimized for an average GC-content of 43% throughout the whole genome (although whether we do this or not does not significantly affect the results). The alignments of each copy of a RE with its respective master sequence were subsequently used to identify nucleotide substitutions, insertions, and deletions.

The subsequent analysis only takes Alu elements into account (Britten et al. 1988; Jurka and Smith 1988; Kapitonov and Jurka 1996). These are short interspersed nucleic elements (SINEs) that have been inserted into the human genome after the human-mouse split. Further we restrict our analysis only to the following 8 Alu sub-families: AluJo, AluJb, AluSx, AluSq, AluSp, AluSg, AluSc, and AluY. These Alu families are those that are most often represented in the human genome. The oldest such elements diverged up to about 20% from their consensus sequence. Nevertheless, they are easy to identify in the human genome and to classify into different sub-families associated with different burst of Alu into the human genome. We exclude the poly-A tail of the Alus and the A-rich linker between the two Alu-monomers from our analysis. Due to multiple repetition of a single base these regions are thought to evolve differently due to replicase stutter and a simple nucleotide substitution model would not properly describe the evolution of poly-A tails. In our previous analysis (Arndt et al. 2003b) we show that very similar substitution patterns can be obtained from other recently inserted REs such as long interspersed nucleic elements (LINEs). We convinced ourselves that this holds also true for the current analysis. However, LINE elements are not as prevalent as Alu repeats and the analysis cannot be performed with the same spatial resolution.

**Substitution model**

We assume that each Alu sub-family has been inserted during a short period of time during human evolution. After insertion each single copy of Alu evolved independently and accumulated base substitutions with respect to the neutral background substitution pattern in the respective region.

In total there are 12 distinct substitution processes of a single nucleotide by another. Four of them are so-called transitions that interchange a purine with a purine or a pyrimidine with a pyrimidine. The remaining eight processes are the so-called transversions that interchange a purine with a pyrimidine or vice versa. Considering only substitutions along neutrally evolving DNA however, we cannot distinguish the two strands of the DNA and therefore the substitution rates are reverse complement symmetric, e.g. the rate for a substitution C→A is equal to the rate for a substitution T→G (in the following we will denote this rate by $r_{C:G \to A:T}$).

In vertebrates it is also important to consider the neighbor-dependent base substitution of cytosine in CpG resulting in TpG or CpA. This process is triggered by the methylation and subsequent deamination of cytosine in CpG pairs (Coulondre et al. 1978; Razin and Riggs 1980). It is commonly (and erroneously) assumed that the CpG-based transition only affects CpG dinucleotides, and that one can learn about all the other substitution processes by excluding CpG sites from the analysis (see for example (Hardison et al. 2003)) This is based implicitly on the view that the CpG process is a small perturbation to the neighbor-independent substitution model described above. However, the CpG-based transition is actually the predominant substitution process in mammals with a rate about 50x higher than that of a transversion (Arndt et al. 2003b; Hess et al. 1994). The importance of this process have also been discussed elsewhere (Arndt et al. 2003a; Duret and Galtier 2000; Fryxell and Zuckerkandl 2000) We include this process explicitly in our model to analyze substitution pattern in the human lineage. The model itself is parameterized by the substitution rates and the length of the time span, $dt$, the respective substitution processes acted upon the sequence, which in our case is the time after insertion of a particular RE, $T$. We have the freedom to rescale time and actually measure it in units of $T$. In this case, the time span is $dt = 1$ for every subfamily and the substitution rates are equal to the substitution frequencies giving the number of base substitutions per nucleotide after the insertion of the RE. (The total number of nucleotide substitutions of one kind during this time equals to the product of the corresponding frequency and the numbers of nucleotides in the respective ancestral state.) The seven parameters of our model are: 4 transversion frequencies ($r_{C:G \to A:T}$, $r_{A:T \to T:A}$, $r_{C:G \to G:C}$, $r_{A:T\to C:G}$), 2 transition frequencies $r_{C:G\to T:A}$, $r_{T:A\to C:G}$) and the frequency of CpG deamination, $r_{CpG}$. We will denote this set of 7 quantities by $\{r\}$.

In order to facilitate the subsequent maximum likelihood analysis (see below) we need to compute the probability, $P_{\{r\}}(.\beta.|\alpha_1\alpha_2\alpha_3)$, that the base $\alpha_2$ flanked by $\alpha_1$ to the left and by $\alpha_3$ to the right, changes into the base $\beta$ for given substitution frequencies $\{r\}$. This probability can easily calculated by numerically solving the time evolution of the probability to find three bases $p(\alpha\beta\gamma;t)$ at time $t$, which is given by the Master equation and can be written into the following set of differential equations:

$$
\begin{aligned}
\frac{\partial}{\partial t} p(\alpha\beta\gamma;t) = &\sum_{\varepsilon\in\{A,C,G,T\}} [r_{\varepsilon\to\alpha}\, p(\varepsilon\beta\gamma;t) + r_{\varepsilon\to b}\, p(\alpha\varepsilon\gamma;t) + r_{\varepsilon\to c}\, p(\alpha\beta\varepsilon;t)] \\
&+ r_{CpG}(\delta_{CA,\alpha\beta} + \delta_{TG,\alpha\beta})\, p(\mathrm{CG}\gamma;t) \\
&+ r_{CpG}(\delta_{CA,\beta\gamma} + \delta_{TG,\beta\gamma})\, p(\alpha\mathrm{CG};\mathrm{t}) \\
&- 2r_{CpG}(\delta_{CG,\alpha\beta} + \delta_{CG,\beta\gamma})\, p(\alpha\beta\gamma;t)
\end{aligned}
$$

where the numbers $r_{\alpha\to\alpha}$ are defined as $r_{\alpha\to\alpha} = -\sum_{\beta\neq\alpha} r_{\alpha\to\beta}$ and $\delta_{\alpha\beta,\gamma\delta}$ is the Kronecker delta defined as

$$
\delta_{\alpha\beta,\gamma\delta} = \begin{cases} 1 & \text{if } \alpha=\gamma \text{ and } \beta=\delta \\ 0 & \text{otherwise} \end{cases}
$$

The first three terms describe single nucleotide substitutions on the three sites whereas the last 4 terms represent the neighbor dependent CpG based transition for CpG's on sites on (1,2) or (2,3). The system is initialized at $t = 0$ with the presence of the three bases $\alpha_1\alpha_2\alpha_3$:

$$
p(abc;t=0) = \begin{cases} 1 & \text{if } (abc) = (\alpha_1\alpha_2\alpha_3) \\ 0 & \text{otherwise} \end{cases}
$$

After numerically iterating the above differential equations using for example the Runge-Kutta method (Press et al. 1992) we get the above transition probability by

$$
P_{\{r\}}(.\beta.|\alpha_1\alpha_2\alpha_3) = \sum_{\beta_1\beta_3} p(\beta_1\beta\,\beta_3;t=1)
$$

The above iteration has to be carried out for all possible 64 combinations of bases $\alpha_1\alpha_2\alpha_3$ to get all 256 possible probabilities $P_{\{r\}}(.\beta.|\alpha_1\alpha_2\alpha_3)$. Note that the above set of differential equations can easily be extended to describe the evolution of sequences with $L > 3$ nucleotides. In this case one gets $4^L$ equations. However, for our subsequent maximum likelihood analysis we just need the time evolution of three consecutive nucleotides, with neighbor dependent processes only at positions (1,2) and (2,3).

### Estimation of substitution frequencies

One can estimate the above mentioned substitution frequencies $\{r\}$ by comparing a pair of master $\vec{\alpha} = \alpha_1\alpha_2 \ldots \alpha_L$ and daughter sequence $\vec{\beta} = \beta_1\beta_2 \ldots \beta_L$, where the daughter sequence represents the state of the master sequence after the substitution processes acted upon it for some time. The log likelihood that the sequence $\vec{\beta}$ evolved from the master sequence $\vec{\alpha}$ under a given substitution model parameterized by the substitution frequencies $\{r\}$ is in very good approximation (Arndt 2004; Arndt et al. 2003a) given by

$$\log L = \sum_{i=2}^{L-1} \log P_{\{r\}}(.\beta_i. \mid \alpha_{i-1}\alpha_i\alpha_{i+1})$$

To get estimates for the substitution frequencies for given $\vec{\alpha}$ and $\vec{\beta}$ we maximize the above likelihood by adjusting the substitution frequencies. This can easily be done using Powell's method (Press et al. 1992) while taking care of boundary conditions (Box 1966), i.e. the positivity of the substitution frequencies. Due to the stochastic nature of the mutation process the estimates of substitution frequencies will be inaccurate within some limits. Supplying more sequence data to the algorithm reduces the error in the frequency estimates. To make sure that the expected uncertainties of estimates from finite sequence data are below 5% and do not influence our results, we estimated frequencies only if more than 5000 bp of aligned sequence data was available. More details on this algorithm, especially a discussion on the expected uncertainties is presented elsewhere (Arndt 2004).

### Estimation of substitution frequencies using repetitive elements

Various repetitive elements that were introduced into the human genome in high copy number serve as pairs of master and daughter sequences. These repetitive elements can be identified and aligned to the ancestral master sequence using the RepeatMasker. The alignments may also contain gaps representing insertions or deletions of bases. In our analysis, we disregard all sites $i$ in the above expression for the likelihood where either one of the symbols $\alpha_{i-1}, \alpha_i, \alpha_{i+1}, \beta_i$ is representing a gap. Subsequently we can estimate genome wide substitution frequencies $\{r\}^F$ for each sub-family of RE's $F$ by taking all identified copies of the RE into account. Further we can estimate regional substitution frequencies $\{r\}^{F,x}$ by taking only those copies of a RE into account which are found in the genomic region $x$. In our case these regions are non-overlapping intervals of 1 Mbp along the chromosomes. The age of a particular sub-family of repeats, $a^F$ , can be defined as the genome wide average transversion frequency. It has already been shown by (Arndt et al. 2003b) that the transition frequencies show a remarkably good linear correlation with the transversions frequencies. To minimize the statistical error in this quantity we take the sum of the transition and transversions frequencies to define the age of a RE. We also tried other definitions of the age, e.g. the sum of the two GC-neutral transversions rates, $r_{A:T\to T:A}+r_{C:G\to G:C}$, however the final results do not change significantly.

**Estimation of a RE averaged regional substitution pattern**

To reduce the stochastic error in regional substitution pattern we can estimate a relative substitution pattern $\{r\}^x$ by considering the weighted sum for each of the seven processes relative to the age of the RE:

$$r^x_{process} = \frac{\sum_F (r^{F,x}_{process} / a^F)(a^F)^2}{\sum_F (a^F)^2} = \frac{\sum_F r^{F,x}_{process}\, a^F}{\sum_F (a^F)^2}$$

Algebraically this expression represents also the slope of a fitted straight line through the origin (using least squares, see Figures 1A and B). The resulting $r^x_{process}$ represents the regional substitution frequency relative to the clock set by the average transversions frequency. Some families are represented only little (less than 5000 bp) in various regions. In this case we exclude such families from the above expression as a safeguard against inaccuracies in the estimation process of the substitution frequencies for these families. If none of the considered REs is sufficiently represented the whole region is excluded from further analysis.

**Other genomic features**

Besides measuring substitution frequencies in the various regions we can also measure other quantities along the human chromosomes as the base composition, the GC-content, or the density of exons. Besides the genomic GC-content one can also define the GC-content of non-featured sequence, i.e. the GC-content of the sequence after removing all featured sequences like exons, repetitive elements, or CpG islands (as defined and annotated in the ensembl database (Hubbard et al. 2002). However, the latter quantity is highly correlated (R=0.88) with the genomic GC-content of all the bases in the 1 Mbp region and we restricted our discussion to the genomic GC-content only. The amount of genic sequence in each region is measured by the exon density, which is defined to be the fraction of exonic nucleotides. The annotation of exons and the information about G-banding densities where taken from the databases at www.ensembl.org (Hubbard et al. 2002). The recombination rates are sex averaged recombination rates taken from (Kong et al. 2002) after remapping the marker position onto the newer genome assembly we use for our analysis.

### Correlation coefficients

Because the observed substitution frequencies and other genomic features are not Gaussian distributed we always use non-parametric tests and report Kendall's tau for pair correlations. For the computation of partial correlations $\tau_{AB,C}$ of two quantities A and B controlling for a third one C we use (Kritzer 1980)

$$\tau_{AB,C} = \frac{\tau_{AB} - \tau_{AC}\tau_{BC}}{\sqrt{1-\tau_{AC}^2}\sqrt{1-\tau_{BC}^2}}$$

where $\tau_{AB}$ are Kendall's tau correlation coefficients of the two quantities A and B.

### Fitting of 4th order polynomials

To quantify the non-linear behavior in the dependence of the seven substitution frequencies on the GC-content $f_{GC}$, we fitted 4th order polynomials to scatter (Figure 5). $f_{GC}$ These polynomials have also been used to calculate the residual substitution frequencies. The coefficients of the polynomials

$$r(f_{GC}) = a_0 + a_1 f_{GC} + a_2 f_{GC}{}^2 + a_3 f_{GC}{}^3 + a_4 f_{GC}{}^4$$

have been computed using standard methods (Press et al. 1992) and their values are given in Table 4.

### Spatial correlations of hotspots

To judge whether the regions with high substitution frequency cluster together or tend to be randomly distributed we mark each region with "+1" if the substitution frequency under consideration is higher than its autosomal median and with "-1" if it is lower than the median. Apparently, half of the region will be marked "+1" and the other half "-1". In Figure 4B we plotted the length distributions of consecutive regions marked with "+1". If the two marks "+1" and "-1" were distributed randomly, the length distribution is given by $N_0 / 2^{n+2}$ where $n$ is the length of run of "+1" and $N_0$ is the total number of regions. This neutral expectation is also given in both Figure 4B.

## *Results*

### Chromosomal profiles with all seven rates

As described in the Methods section we utilized repetitive Alu sequences to measure nucleotide substitution frequencies (corrected for back- and multiple- substitutions) in windows of 1 Mbp along the human chromosomes. In our analysis of substitution patterns, we distinguish 4 transversion processes, 2 transition processes, and the CpG-based deamination process of cytosine transition (CpG→TpG/CpA). In Fig. 1A and B, we provide details of the analysis of these substitution processes for two 1 Mbp regions taken from Chromosome 1. For each 1 Mbp region the above 7 corrected substitution frequencies are computed for each of the 8 subfamilies of Alu's found in that region. AluY is the youngest family of Alu's, and accordingly, the substitution frequencies obtained are the lowest. AluJo is the one of the oldest families, and are found to have the highest substitution frequencies. To combine the results obtained from the different Alu subfamilies, we plot the 7 substitution frequencies against the genomic-averaged transversion frequency for the respective Alu subfamily. We find that the data are nearly linear for each of the 7 substitution process, suggesting that it is valid to use a simple linear function to characterize the relative substitution frequencies experienced by the region throughout the time period (nearly 90 Myr) during which these Alu elements were deposited (see Methods for details.)

The slopes obtained from the two top panels in Fig. 1 clearly show that the substitution processes are quite different in these two regions. To illustrate the regional differences across the chromosome in more detail, we plot in Fig. 1C the regional average of the 4 transversion frequencies for 3 distinct Alu families (AluY, AluSx, and AluJo) along the entire length of Chromosome 1. The figure clearly shows regional variations across the chromosome, with certain "hot spots" (e.g., at 185 Mb) where the obtained frequencies are ~50% higher than those at typical regions. Importantly, we observe that the regional variations are highly correlated across the Alu families (pairwise correlation coefficients are greater than 0.55 for the 3 quantities mentioned above). This justifies the use of the slopes of Figs. 1A and B as the time-average substitution frequencies for each region. It also reduces the statistical error in the regional estimates of substitution frequencies, thereby allowing us to sharpen the spatial resolution to the 1 Mbp scale.

Profiles of the seven time-averaged substitution frequencies along all human chromosomes can be found in the Supplementary Material, which is available at the publishers web site.. All substitution frequencies are rescaled such that the average transversion frequency averaged across the whole genome occurs at a rate of 1 per bp (see Methods). Relative to transversions the transition T:A to C:G is observed about 3 times, and the transition C:G to T:A about 5 times more frequently (Arndt et al. 2003b; Lander et al. 2001). The fastest substitution process is the CpG-based transition, which can be observed about 50 times more frequently than an average transversion (Arndt et al. 2003b). For comparison of the substitution frequencies with other genomic features we present more details for chromosome 1 in Fig. 3. We show the G-banding pattern (Fig. 3A) and the genomic GC-content, exon density (i.e. the fraction of exonic nucleotides), and recombination rate (Fig. 3B) on the same spatial scale. Similar figures for the other 23 chromosomes show similar behaviors (see Figures S1-S24 in the Supplementary Material). In this paper we will exclude the two sex chromosomes, X and Y, from further analysis and discuss only autosomal substitution pattern.

Overall, the profiles for the seven frequencies are fairly smooth. However there are several distinct regions on all chromosomes where some of the frequencies are simultaneously elevated, for example on chromosome 1 in the regions from 100 – 110 Mbp, 185 – 190 Mbp, and other smaller regions (Fig. 3C). In these regions the frequencies of the three processes C:G→G:C, C:G→A:T, and C:G→T:A are elevated by at least 50% and clearly stand out of the statistical noise. This anomalous behavior of the three processes which exchange the C:G base pair by any other (and collectively referred to below as C:G→X:Y) can also be seen in the distribution of the substitution frequencies. As shown in Fig. 4A, the distributions for these 3 processes are highly asymmetric (the skewness of the latter distributions is 1.3 and higher indicating broad tails of the distribution towards higher frequencies). In Fig. 4B we show also the length distribution of adjacent regions with substitution frequencies above the respective median frequency. Comparing the latter distributions to that of a spatially randomized version of the genome (the thick black line in Fig. 4B) we find that those regions that accumulated more than the median substitutions extend beyond that what is expected by chance.

**Correlations of the substitution rates and several genomic parameters**

To learn more about those substitution pattern and their possible causes and consequences we plotted several other genomic features such as the G-banding pattern, the GC-content, recombination rate, and the exon density along chromosome 1 in Fig. 3A and B. Visual inspection reveals apparent correlations of the three C:G exchanging processes with these features. Regions that show more than average substitutions tend to have darker G-bands, are AT rich, and have fewer exons. We will confirm these observations in further analyses below. To begin our exploration of the underlying causes of rate variation we have calculated a set of pairwise correlations (using Kendall's rank correlations approach) between these genomic parameters and all of the seven rates (Table 1).

Among one another the frequencies for the three substitutions away from C:G base pairs (C:G→X:Y) show the strongest correlations. All correlations with the frequencies of the processes that exchange A:T base pairs (A:T→X:Y) are much weaker ($|\tau| \leq 0.15$). The genomic parameters also show complex patterns of correlations with each other. In particular, we confirm previous results (Lander et al. 2001) showing that exon density is strongly positively correlated with GC content (Kendall Rank Correlation, $\tau = 0.536$, $P < 0.0001$). We also confirm (Cheung et al. 2001; Furey and Haussler 2003; Mouchiroud et al. 1991; Saccone et al. 1993) that the dark G-bands are GC- and exon-poor (Kendall Rank Correlations, $\tau = -0.509$ with GC content and $\tau = -.353$ with exon density; for both $P < 0.0001$). Among the substitution rates, the three C:G→X:Y substitution rates show the strongest (negative) correlations with the GC-content and exon density, while having positive correlations with the G-staining density.

We want to better understand these relationships between correlated quantities. To find out whether one factor is the cause or the consequence of another factor, or whether the correlation between two factors is actually mediated by a third factor, we computed partial rank correlations of the seven substitution frequencies with genomic features and controlling for the other features in turn (Table 2A). The relationship between GC-content and the C:G→X:Y substitution rates appears to be the most robust. When controlling for other features the partial correlations with the GC-content stay higher than partial correlations in all other combinations.

To understand the form of the relationship between GC content and the substitution frequencies, we plotted these relationships in Fig. 5. Several patterns are apparent from these scattergrams. First, there appears to be substantially more scatter in the rates in the low GC regions for all the rates. The F-tests of variance confirm that the variance is greater in the regions of low GC content (GC% is less that the genome-wide median value of 41%) than in the regions of high GC content (GC% is larger than 41%) for all the rates ($P < 0.0001$ in all cases, see Table S1 in the Supplementary Material). The largest difference was observed for transversions C:G→A:T and C:G→G:C (respectively 5.4 and 3.6 times higher variance in the regions of low GC). These two transversions also show the most dramatic non-linear increase in the rates in the region of low GC (Fig. 5).

The nonlinearity of the relationship between GC content and at least some of the substitution rates suggests that it might be more appropriate to calculate correlations of the substitution rates and genomic features in the GC-poor and GC-rich regions separately. Table 2B shows all of the calculated correlations and partial correlations in the GC-poor regions (GC content is less than the median value of 41% and Table 2C shows the same correlations in the GC-rich regions (GC content is higher than 41%). Comparing the latter two tables with Table 2A, it becomes clear that the correlations between the substitution frequencies and genomic features are much stronger in the regions with low GC-content and are much weaker in regions with high GC-content (Table 2B).

**Understanding GC-independent rate variation**

To better understand GC-independent effects on the rates of substitution, we statistically removed effects of GC-content by fitting the relationship of each rate and the GC-content with a 4$^{th}$ order polynomial curve (the curves in Fig. 5). (The fitted polynomials can be found in the Methods section.) Profiles of the residuals of these polynomial regressions relative to the corresponding substitution frequency for the three C:G→X:Y processes are shown in Fig. 3D. As in the panel above (Fig. 3D) one can clearly find regions where these three relative residuals are elevated (for instance regions 45 - 50 and 100 – 110 Mbp). However, some regions with elevated substitution frequencies do not show elevated residuals (such as, for instance, region 185 – 190 Mbp). In the latter cases, the higher substitution frequencies are in correspondence to the particularly low GC-content in those regions. As in the case of the substitution frequencies, the distributions of relative residuals are also not symmetric and have a tail for high relative residual substitution frequencies (see Fig. S25A in the Supplementary Material) highlighting that there are more regions with unusually elevated substitution frequencies than there are with unusually low frequencies after accounting for the different GC-content in the respective regions. Again, these regions of elevated residuals are longer than expected after comparison with randomly shuffled version of the genomic sequence (Fig. S25B).

To analyze the residual variations further, we report the correlations of the residuals for all of the rates with each other in Table 3. Correlations with various genomic parameters are presented in TableS2 in the Supplementary Material. All correlations are significantly reduced compared to those reported in Tables 1 and 2, consistently with GC-content being the strongest predictor of substitution rates. However, the correlation between residuals of the two transversions C:G→G:C and C:G→A:T is still very high (Kendall's rank correlation, $\tau = 0.4$), indicating that these processes are coupled. As it can be seen in Table S2C there is also significant correlation between the regional exon density and the residual substitution frequencies in regions with low GC-content. These correlations are much lower when considering the whole data set (Table S2A) or regions with high GC-content (Table S2B). Residuals for the C:G→X:Y transversions are on average substantially higher in gene poor regions as can be seen in Figure S26 where we plotted the averaged residuals in regions with different exon density.

**Telomere effects**

It is possible that the location on the chromosome relative to the structures such as telomeres or centromeres has an influence on nucleotide substitutions. In particular there are reasons to expect differences in the patterns of substitution near telomeres as they have been shown to have systematically higher GC-content compared to other regions of the genome (Bernardi 2000). We therefore partitioned our data with respect to the distance to the nearest telomere averaged over all autosomes. In Fig. 6A we show the values of the GC-content and exon density in relation to the autosomal means. We observe that indeed the GC-content is about 20% higher in telomeric regions and reaches its autosomal mean at a distance of about 15 Mbp. The exon density is more than twice as high at the very tips of the chromosomes.

The relative substitution frequencies are depicted in Fig. 6B. Shown are again the relative frequencies relative to the autosomal means. Clearly, the GC-enriching processes are enhanced at the chromosomal tips (about 20 Mbp from the end). On the other hand, the GC-depleting processes are suppressed, while the GC-neural processes stay at about their mean value.

Based on our analysis of all autosomal regions, we would expect a different behavior at the GC-rich telomeres. In particular, although the observed reduction in the C:G→X:Y rates is not unexpected, the increase of the rates of the reverse substitutions (A:T→X:Y) is not predicted by the higher GC-content (Fig. 5) or the higher exon density (Fig. S26). To quantify these expectations we plotted the residuals of the rates after accounting for the average dependence of the rates on GC-content. This analysis shows that the residuals of the of A:T→X:Y are indeed unusually high (Fig. 6C). These analyses show that telomeric regions do exhibit a different substitution pattern from the rest of the autosomal regions. We can further state that this pattern only extends at most 15 Mbp from the tips into the chromosomes. Beyond this scale the substitution pattern is not significantly different from the autosomal mean, apart from those regions with simultaneously enhanced C:G→X:Y rates as discussed previously.

## Discussion

**1 Mbp scale map of substitution rates**

In this study, we provide the first map of substitutional patterns across human autosomes at a 1Mbp resolution. Such a high resolution is made possible by our use of a very large number of Alu elements inserted into the human genome since the mammalian radiation. Beyond the main advantage of the very large amount of sequence data, the use of Alu sequences provides us with the ability to estimate substitutional patterns within a consistent sequence context while varying the location of the sequence across the genome or varying the amount of time the sequence has been undergoing evolution.

The inference of substitutional patterns based on ancient insertions of Alu sequences is not a very straightforward task. Ancestral Alu sequences have a high GC and a high CpG content (Britten et al. 1988; Jurka and Smith 1988) An exceptionally fast rate of C to T transition at methylated cytosine commonly found in CpG pairs (CpG→TpG/CpA transition) presents a great challenge to all studies of substitutions in the mammalian genomes. The standard approach of excluding CpG sites is deeply unsatisfying for two reasons. First of all, it excludes a very powerful substitution process from consideration. What is even worse, the exclusion of ancestral CpG sites fails to eliminate the effect of CpG→TpG/CpA transitions on other substitution processes. This is because new, non-ancestral CpG sites are generated in the course of the sequence change and their fast resolution to TpG/CpA pairs biases estimates of all rates (Arndt et al. 2003b).

To avoid these problems here we use a maximum likelihood approach pioneered by Arndt, Burge and Hwa (2003a). This approach explicitly incorporates CpG→TpG/CpA process and in principle is able to recover true rates of the seven distinct substitutions (six single base pair substitution rates and the rate of the CpG→TpG/CpA substitution) significantly past the time of the saturation of the CpG→TpG/CpA process (Arndt et al. 2003b).The success of the approach depends on the amount of sequence information; to ensure that the estimation of substitution frequencies for a particular subfamily of Alu's is not dominated by statistical noise we require that within the region under consideration at least 5000 bp are attributed to this Alu subfamily.

We have estimated the rates of substitution by combining information from several different Alu families. This has allowed us to reduce the amount of statistical noise by having more sequence information for any given 1 Mbp. This also allowed us to test whether the estimates are internally consistent and also to see whether the location of hotspots has been consistent through time. Fig. 1 depicts the estimates derived from individual families to demonstrate that they do give very consistent results. Overall there is a positive correlation for all the rates derived from individual Alu families, which is strongest between those substitutions that show more variations along the chromosomes, i.e. the substitutions C:G to X:Y. This suggests further that our estimates are consistent and also that the relative substitution rates in any particular 1 Mbp region tend to be stable over the studied time of ~90 Myr.

**Strong and large-scale dependence of the C:G to X:Y rates on the GC content**

Our analysis revealed substantial, ~2-fold variation in the substitution rates across the human genome. Intriguingly, both visual inspection and the statistical analysis show that the distribution for all rates is skewed towards higher rates. This indicates that more regions of the genome are hotspots of substitution and very few regions are coldspots. For example, if we only look at the rate of C:G→A:T transversions, then there are 101 regions with more than 1.5-fold higher and none with less then 0.5-fold the genome-average rate. If we relax the condition to 1.3-fold versus 0.7-fold of the genome average, then the ratio becomes 312 to 10. It also appears that many of the hotspots extend substantially beyond the 1 Mbp region. For example there are 23 regions of 10 Mbp or longer that have substantially elevated rates of the C:G→A:T transversions compared to the expectation of only 1 such region if the hotspots were distributed independently of each other (Fig. 4).

To gain insight into the causal forces that bring about this variation in the substitution rates, we conducted a comprehensive analysis of correlations among all of the rates with each other and also with various genomic features, such as GC content, exon density, recombination rate and the distance to telomeres. To our surprise the pairwise correlations among the substitution rates are much stronger for C:G to X:Y substitutions (including the two transversions and both the methylation-dependent and methylation-independent transitions) than for any of the A:T to X:Y substitutions. We also found that the rates of C:G to X:Y substitutions correlate strongly and negatively with the GC-content and exon density and strongly but positively with the G-bands in the chromosomes. We found much weaker correlations with the rates of recombination. However, it is possible that the current estimates of recombination rate are too crude to detect the existing correlations.

In the course of this analysis we confirmed that GC-content, exon density and the presence of G-bands all correlate with one another (Cheung et al. 2001; Furey and Haussler 2003). High GC-content regions tend to be exon rich and stain lightly with Giemsa dye (e.g. tend not to be labeled as G-bands). Thus to unravel the underlying causes of the substitution rate variability we conducted all possible pairwise partial correlations (Table 2A). This analysis revealed that GC-content is the main predictor of the substitution rates (again primarily of the C:G to X:Y rates, since the A:T to X:Y rates vary very little). Controlling for GC content substantially reduces correlations among the substitution rates and exon density or the presence of G-bands, whereas correlations of substitution rates with GC-content stay high when we control statistically for exon density and G-banding.

Plots of the substitution frequencies versus GC-content in Fig. 5 further distinguishes the behaviors of the two C:G–based transversions and the C:G to T:A transition. The C:G-based transversion frequencies exhibit sharp increases in regions of low GC content, essentially *doubling* from regions with 40% GC-content to regions with 35% GC-content. Although the frequency of C:G to T:A transition is also higher in the low GC areas, this increase is much more gradual and appears to be constant throughout all regions of GC-content. E.g., there is only ~10% increase of the substitution frequency from 40% to 35% GC-content. These features are also reflected in the partial correlations conducted separately for the high and low GC regions (Tables 2B and 2C), confirming that there is a strong GC-dependence of the C:G to X:Y frequencies in the low GC regions, but that it is much weaker (especially for the C:G–based transversions) in regions of high GC content. It is possible that there is a separate process governing C:G-based transversions in the low GC regions.

**GC-independence of many hotspots of C:G to X:Y transversions**

The strong dependence of the C:G to X:Y substitutions on GC content raised the possibility that many of the hotspots of C:G to X:Y substitutions that we identified initially are due to the overabundance and the clustering of the regions of particularly low GC content. To investigate this possibility we decided to investigate the GC-independent behavior of all of the substitution rates. To achieve this we first found the best fit of a 4th order polynomial function between the rate of a particular substitution and the GC content of the genomic region (Fig. 5 and Table 4). We then interpreted the deviations of the observed rates compared to the expected rate based on this polynomial fit as the GC-independent substitution rate variation.

The GC-independent rates of substitution show some interesting behaviors (Tables 3 and S2, and Fig. 3D). Given that most of the correlations are statistically mediated through GC content, the GC-independent rates, as expected, show very little dependence not only on GC content, but also on other genomic features. Surprisingly, however, we still observed very strong positive correlations of the GC-independent rates of the two C:G to X:Y transversions with each other (Kendall's correlation, $\tau = 0.37$). This suggests that the rates of the C:G to X:Y transversions are *systematically* affected by a common causal factor that is independent of the GC-content of surrounding sequences. This possibility is also consistent with the similarly curvilinear relationships between each C:G to X:Y transversion and GC content and the very large amount of variability in the rates of these two transversions in the low GC regions (Table S1). It is also interesting that the distribution of GC-independent residuals of C:G to X:Y transversions is also skewed having more weight for higher residuals (Fig. S25A). Hence, even after accounting for GC there is still a higher proportion of hotspots compared to coldspots.

Intriguingly, it appears that part of the reason for these high rates is the acceleration of both C:G to X:Y transversions in the most gene-poor regions. See Figure S26 where we plot the GC-independent residuals of all of the rates versus exon density of each 1 Mbp window and only C:G to X:Y rates show a systematic dependence. Note that, surprisingly, we do not observe higher rates of CpG to TpG/CpA methylation-dependent transitions given that methylation rates are thought to be higher in gene poor regions (Rabinowicz et al. 2003; Yoder et al. 1997). It is possible that variability of methylation rates takes place at smaller scales than 1 Mbp level of resolution in the present analysis. We do, however, observe substantially higher rates of C:G to X:Y transversions.

Mapping of the GC-independent rates on the chromosomes revealed that some of the hotspots of C:G transversions that we identified initially (Fig. 3C) are indeed due to the existence of regions of particularly low GC content. For example, two hotspots located between 50 – 80 Mbp on chromosome 1 (Fig. 3C) disappear after accounting for GC content (Fig. 3D). However, some of the largest hotspots, such as the one located between 100 and 110 Mbp on the chromosome 1, are still present after accounting for the systematic variation of the rates with GC content. Visual inspection (compare Figs. 3C and 3D) suggests that approximately 50% of the hotspots of C:G transversions remain after accounting for the GC-content. To get a more quantitative idea about all autosomal regions we fitted Gaussian distributions to the distributions of substitution frequencies as shown in Fig. 3C). The respective values for the mean and standard deviation are given in Table 5 together with the expected number of 1 Mbp regions deviating more than one standard deviation form the mean. In the same table we give also report the actual number of 1 Mbp regions where the respective substitution frequency deviates more than one standard deviation from the mean. Clearly, we observe more than expected hot regions for the two C:G to X:Y transversion. However, about one third of these regions drop out if we also demand that the residuals deviate more than 1 standard deviation from their mean.

**Unusual patterns of substitution at telomeres**

A striking finding of our study is the distinct pattern of substitution close to the telomeres of all chromosomes. Telomeres have been known to have an elevated GC content, but there are many other regions in the genome with equally high GC contents. Telomeres also have an unusually high density of genes (Bernardi 2000). Based on the high GC content we would expect somewhat diminished rates of C:G to X:Y substitutions but no real change in the A:T to X:Y substitutions. The high gene density should also suppress the rates of both C:G to X:Y transversions.

The observed pattern is quite different. The C:G to X:Y rates are indeed lower than genome average. However, surprisingly the rates of A:T to C:G transversion and A:T to G:C transition are sharply elevated. Interestingly, this cannot be explained by a simple elevation of mutability of A:T pairs given that A:T to T:A rates do not appear to change near telomeres. Only the rates of those A:T to X:Y substitutions that increase GC content are elevated. Similarly, C:G to G:C rates are not lower near telomeres, whereas all rates that change a C:G pair into an A:T pair are reduced.

Also, intriguingly, the telomere effect on the genomic properties such as GC-content and exon density extends approximately 10, at most 20 Mbp. Telomere-specific patterns of substitution also appear to extend for the same 10-20 Mbp range from the telomere. The coincidence in the ranges is suggestive of the causal relationship among the patterns of substitution and the genomic properties near telomeres.

**Mutation or selection?**

The variation in substitution rates in principle could be due to differences either in mutation rates or in the probabilities of fixation (brought about by either selective differences or via DNA repair biases resulting in biased gene conversion (BGC) of different nucleotides across the genome. It is often very difficult to distinguish between these two possibilities.

One interesting pattern in our data is the sharp variation in the rates of the C:G→A or T transversions but very muted variation in the rates of reverse transversions (A:T→C or G). This pattern can be easily explained by the variation in the rates of mutation if, for whatever reason, C:G pairs are more likely to undergo transversion mutations in the regions of low GC content. Intuitively it is somewhat harder to explain this by differences in the probabilities of fixation, given that the fixation of A:T pairs and C:G pairs should be correlated. Indeed, both of the substitutions have to go through a stage where C:G vs. A:T polymorphism is segregating at a particular site. Thus the change of fixation probability of A:T pairs should have an equal in magnitude and opposite in direction effect on the probability of fixation for C:G pairs. To the extent that differences in the rate of C:G to A or T substitutions are brought about by natural selection or BGC, we would expect to find the opposite effect on the rates of the reverse,

A:T to C or G substitutions. Instead we see almost no variation in the later rates. We also want to mention a recent study that reports correlations between the stationary GC-content calculated from neutral substitution patterns and the rate of crossovers (Meunier and Duret 2004), and concludes that recombination drives the base composition of genomes. The full treatment of these questions will be presented elsewhere.

In addition we find that there is a strong correlation, and likely a similar underlying cause, for C:G to A:T and C:G to G:C transversions. The first one of them is GC-content depleting, while the second one does not affect the GC content. It is hard to imagine a similar cause at the level of fixation acting on both these substitutions given this difference.

In contrast, the pattern that we observe at telomeres is entirely consistent with either natural selection of BGC promoting a higher GC content at telomeres. The rates of GC-enriching substitutions are elevated while the rates of GC-depleting substitutions are reduced. There is also no effect on the rate of substitutions that do not affect GC-content (C:G to G:C and A:T to T:A).

**Acknowledgement**

TH is supported by the NSF through grant 0211308, 0216576, 0225630. DP is supported by the NSF grant DEB-0317171, the Terman Award, and the Alfred P. Sloan Fellowship in Computational Molecular Biology. PA and DP are grateful to the hospitality of the Center for Theoretical Biological Physics at UCSD where extensive discussions on this research took place.

Cited Literature

## *Tables*

| | A:T→C:G | A:T→T:A | C:G→G:C | C:G→A:T | A:T→G:C | C:G→T:A | CpG→CpA/TpG | GC-content | exon density | recombination rate | G staining density |
|---|---|---|---|---|---|---|---|---|---|---|---|
| A:T→C:G | 1.00 | 0.06 | 0.02 | 0.01 | 0.18 | 0.01 | -0.11 | 0.05 | 0.01 | 0.03 | 0.00 |
| A:T→T:A | | 1.00 | 0.09 | 0.13 | 0.11 | 0.23 | 0.03 | -0.17 | -0.11 | -0.05 | 0.10 |
| C:G→G:C | | | 1.00 | **0.52** | -0.05 | **0.36** | 0.26 | **-0.38** | **-0.36** | -0.11 | 0.24 |
| C:G→A:T | | | | 1.00 | -0.13 | **0.54** | **0.40** | **-0.57** | **-0.49** | -0.14 | **0.34** |
| A:T→G:C | | | | | 1.00 | -0.07 | -0.15 | 0.14 | 0.10 | 0.05 | -0.11 |
| C:G→T:A | | | | | | 1.00 | **0.34** | **-0.57** | **-0.44** | -0.15 | **0.32** |
| CpG→CpA/TpG | | | | | | | 1.00 | **-0.44** | **-0.42** | -0.12 | 0.22 |
| GC-content | | | | | | | | 1.00 | **0.54** | 0.18 | **-0.40** |
| exon density | | | | | | | | | 1.00 | 0.05 | **-0.30** |
| recombination rate | | | | | | | | | | 1.00 | -0.11 |
| G staining density | | | | | | | | | | | 1.00 |

**Table 1.** Correlation coefficients (Kendall tau's) of the substitution frequencies and other genomic features. Correlations with absolute values greater or equal 0.3 are typeset in boldface.

| | GC-content | GC-content cntr exon density | GC-content cntr recombination rate | GC-content cntr G staining density | exon density | exon density cntr GC-content | exon density cntr recombination rate | exon density cntr G staining density | recombination rate | recombination rate cntr GC-content | recombination rate cntr exon density | recombination rate cntr G staining density | G staining density | G staining density cntr GC-content | G staining density cntr exon density | G staining density cntr recombination rate |
|---|---|---|---|---|---|---|---|---|---|---|---|---|---|---|---|---|
| **A** | | | | | | | | | | | | | | | | |
| A:T→C:G | 0.05 | 0.06 | 0.05 | 0.06 | 0.01 | -0.03 | 0.00 | 0.01 | 0.03 | 0.02 | 0.03 | 0.03 | 0.00 | 0.02 | 0.00 | 0.00 |
| A:T→T:A | -0.17 | -0.14 | -0.17 | -0.15 | -0.11 | -0.02 | -0.11 | -0.09 | -0.05 | -0.02 | -0.04 | -0.04 | 0.10 | 0.03 | 0.07 | 0.10 |
| C:G→G:C | **-0.38** | -0.24 | **-0.37** | **-0.32** | **-0.36** | -0.19 | **-0.35** | **-0.31** | -0.11 | -0.04 | -0.09 | -0.08 | 0.24 | 0.10 | 0.15 | 0.23 |
| C:G→A:T | **-0.57** | **-0.42** | **-0.56** | **-0.51** | **-0.49** | -0.27 | **-0.49** | **-0.44** | -0.14 | -0.05 | -0.13 | -0.11 | **0.34** | 0.14 | 0.23 | **0.33** |
| A:T→G:C | 0.14 | 0.11 | 0.14 | 0.11 | 0.10 | 0.03 | 0.10 | 0.07 | 0.05 | 0.03 | 0.05 | 0.04 | -0.11 | -0.06 | -0.08 | -0.11 |
| C:G→T:A | **-0.57** | **-0.45** | **-0.56** | **-0.51** | **-0.44** | -0.19 | **-0.44** | **-0.38** | -0.15 | -0.06 | -0.14 | -0.12 | **0.32** | 0.12 | 0.22 | **0.31** |
| CpG→CpA/TpG | **-0.44** | -0.28 | **-0.43** | **-0.39** | **-0.42** | -0.24 | **-0.42** | **-0.38** | -0.12 | -0.05 | -0.11 | -0.10 | 0.22 | 0.06 | 0.11 | 0.21 |
| **B** | | | | | | | | | | | | | | | | |
| A:T→C:G | 0.15 | 0.16 | 0.15 | 0.14 | 0.01 | -0.05 | 0.02 | 0.01 | 0.07 | 0.06 | 0.07 | 0.07 | -0.05 | -0.02 | -0.05 | -0.05 |
| A:T→T:A | -0.11 | -0.10 | -0.11 | -0.11 | -0.06 | -0.01 | -0.06 | -0.05 | -0.02 | -0.02 | -0.03 | -0.02 | 0.05 | 0.02 | 0.04 | 0.05 |
| C:G→G:C | -0.03 | -0.01 | -0.03 | -0.03 | -0.05 | -0.04 | -0.06 | -0.05 | -0.03 | -0.03 | -0.03 | -0.03 | -0.01 | -0.02 | -0.02 | -0.01 |
| C:G→A:T | -0.29 | -0.23 | -0.29 | -0.27 | -0.21 | -0.12 | -0.22 | -0.21 | -0.06 | -0.05 | -0.08 | -0.06 | 0.10 | 0.05 | 0.08 | 0.10 |
| A:T→G:C | 0.11 | 0.12 | 0.11 | 0.10 | 0.00 | -0.05 | 0.01 | -0.01 | 0.05 | 0.05 | 0.05 | 0.05 | -0.07 | -0.05 | -0.07 | -0.07 |
| C:G→T:A | **-0.40** | **-0.35** | **-0.40** | **-0.39** | -0.24 | -0.10 | -0.25 | -0.23 | -0.08 | -0.07 | -0.11 | -0.08 | 0.13 | 0.06 | 0.11 | 0.14 |
| CpG→CpA/TpG | **-0.41** | **-0.31** | **-0.40** | **-0.40** | **-0.35** | -0.23 | **-0.36** | **-0.35** | -0.07 | -0.07 | -0.12 | -0.08 | 0.09 | 0.01 | 0.06 | 0.09 |
| **C** | | | | | | | | | | | | | | | | |
| A:T→C:G | -0.07 | -0.05 | -0.07 | -0.06 | -0.06 | -0.03 | -0.06 | -0.05 | -0.01 | 0.00 | -0.01 | 0.00 | 0.06 | 0.04 | 0.05 | 0.06 |
| A:T→T:A | -0.11 | -0.11 | -0.10 | -0.10 | -0.03 | 0.02 | -0.03 | -0.02 | -0.03 | -0.02 | -0.03 | -0.03 | 0.06 | 0.03 | 0.05 | 0.06 |
| C:G→G:C | **-0.47** | **-0.36** | **-0.47** | **-0.44** | **-0.42** | -0.28 | **-0.42** | **-0.39** | -0.07 | 0.01 | -0.07 | -0.05 | 0.22 | 0.09 | 0.16 | 0.21 |
| C:G→A:T | **-0.57** | **-0.46** | **-0.56** | **-0.53** | **-0.49** | **-0.34** | **-0.49** | **-0.47** | -0.09 | 0.01 | -0.09 | -0.07 | 0.26 | 0.11 | 0.19 | 0.25 |
| A:T→G:C | 0.07 | 0.03 | 0.06 | 0.05 | 0.09 | 0.07 | 0.09 | 0.08 | 0.03 | 0.01 | 0.02 | 0.02 | -0.05 | -0.03 | -0.04 | -0.05 |
| C:G→T:A | **-0.39** | **-0.30** | **-0.38** | **-0.36** | **-0.31** | -0.18 | **-0.31** | -0.29 | -0.09 | -0.02 | -0.08 | -0.07 | 0.19 | 0.09 | 0.14 | 0.18 |
| CpG→CpA/TpG | -0.17 | -0.08 | -0.16 | -0.16 | -0.24 | -0.19 | -0.24 | -0.23 | -0.04 | -0.01 | -0.04 | -0.04 | 0.07 | 0.02 | 0.02 | 0.06 |

**Table 2.** Correlation and partial correlation coefficients of substitution frequencies with various genomic features in (A) all regions of the autosome, (B) in GC-rich regions (GC-content greater than 41%), and (C) GC-poor regions (GC-content less than 41%) of the autosome. Correlations with absolute values greater or equal 0.3 are typeset in boldface.

| | A:T→C:G | A:T→T:A | C:G→G:C | C:G→A:T | A:T→G:C | C:G→T:A | CpG→CpA/TpG | GC-content | exon density | recombination rate | G staining density |
|---|---|---|---|---|---|---|---|---|---|---|---|
| A:T→C:G | 1.00 | 0.07 | -0.03 | 0.01 | 0.17 | 0.07 | -0.09 | 0.00 | -0.04 | 0.03 | 0.02 |
| A:T→T:A | | 1.00 | -0.03 | -0.05 | 0.16 | 0.16 | -0.11 | -0.01 | 0.01 | -0.01 | 0.00 |
| C:G→G:C | | | 1.00 | **0.37** | -0.01 | 0.09 | 0.10 | 0.02 | -0.09 | 0.01 | 0.00 |
| C:G→A:T | | | | 1.00 | -0.07 | 0.18 | 0.16 | 0.02 | -0.12 | 0.01 | 0.02 |
| A:T→G:C | | | | | 1.00 | 0.06 | -0.08 | -0.02 | -0.02 | 0.02 | -0.03 |
| C:G→T:A | | | | | | 1.00 | 0.03 | 0.02 | -0.04 | -0.03 | 0.03 |
| CpG→CpA/TpG | | | | | | | 1.00 | 0.01 | -0.11 | -0.04 | 0.00 |
| GC-content | | | | | | | | 1.00 | **0.54** | 0.18 | **-0.40** |
| exon density | | | | | | | | | 1.00 | 0.05 | **-0.30** |
| recombination rate | | | | | | | | | | 1.00 | -0.11 |
| G staining density | | | | | | | | | | | 1.00 |

**Table 3.** Correlation coefficients (Kendall tau's) of the residual substitution frequencies and other genomic features. Correlations with absolute values greater or equal 0.3 are typeset in boldface.

| | $a_0$ | $a_1$ | $a_2$ | $a_3$ | $a_4$ |
|---|---|---|---|---|---|
| A:T→C:G | 9.310389288 | -81.20476485 | 290.2315796 | -462.89438 | 279.3045503 |
| A:T→T:A | 16.66066957 | -142.7362896 | 482.8107306 | -724.6422854 | 406.0013685 |
| C:G→G:C | 103.4566359 | -865.9111085 | 2741.456674 | -3846.893157 | 2018.678171 |
| C:G→A:T | 165.3930547 | -1396.995937 | 4439.59776 | -6246.304334 | 3279.465165 |
| A:T→G:C | -13.7394868 | 100.4442619 | -171.3649543 | 12.91260921 | 127.6843553 |
| C:G→T:A | 11.5800293 | -27.86943156 | 23.50026482 | 37.90712386 | -57.21062181 |
| CpG→CpA/TpG | -1520.33521 | 13784.91356 | -44830.36973 | 64156.62985 | -34223.33299 |

**Table 4.** Coefficients of the 4th order polynomials $r(f_{GC}) = a_0 + a_1 f_{GC} + a_2 f_{GC}^2 + a_3 f_{GC}^3 + a_4 f_{GC}^4$ used to fit the data shown in Fig. 5.

| | Mean | Standard deviation | Expected number of cold/hot spots | Number of cold regions | Number of normal regions | Number of hot region |
|---|---|---|---|---|---|---|
| A:T→C:G | 0.77 | 0.099 | 534 | 308(308) | 1684 | 579(564) |
| A:T→T:A | 0.81 | 0.098 | 542 | 320(320) | 1651 | 600(600) |
| C:G→G:C | 1.14 | 0.107 | 656 | 94(94) | 1428 | 1049(717) |
| C:G→A:T | 1.02 | 0.109 | 710 | 85(77) | 1297 | 1189(676) |
| A:T→G:C | 3.13 | 0.184 | 539 | 587(518) | 1568 | 416(404) |
| C:G→T:A | 5.07 | 0.352 | 541 | 308(229) | 1777 | 486(435) |
| CpG→CpA/TpG | 48.96 | 3.26 | 513 | 321(237) | 1748 | 502(493) |

**Table 5.** For each of the seven substitution processes we list in the first two columns the mean and standard deviation of a fitted Gaussian to the distribution shown in Fig. 4A. The expected number of cold/hot spots, which deviate more than 1 standard deviation from the mean, is given in the third column. In the last three columns the actual distribution of the 2571 one Mbp regions among the three categories cold, hot (substitution frequency deviates more one standard deviation from the mean), and normal is given. In brackets we give the number of those regions where both the respective substitution frequency and the residual frequency deviate more than one standard deviation in the given respective distributions (Figs. 4A and S25A).

## *Figures*

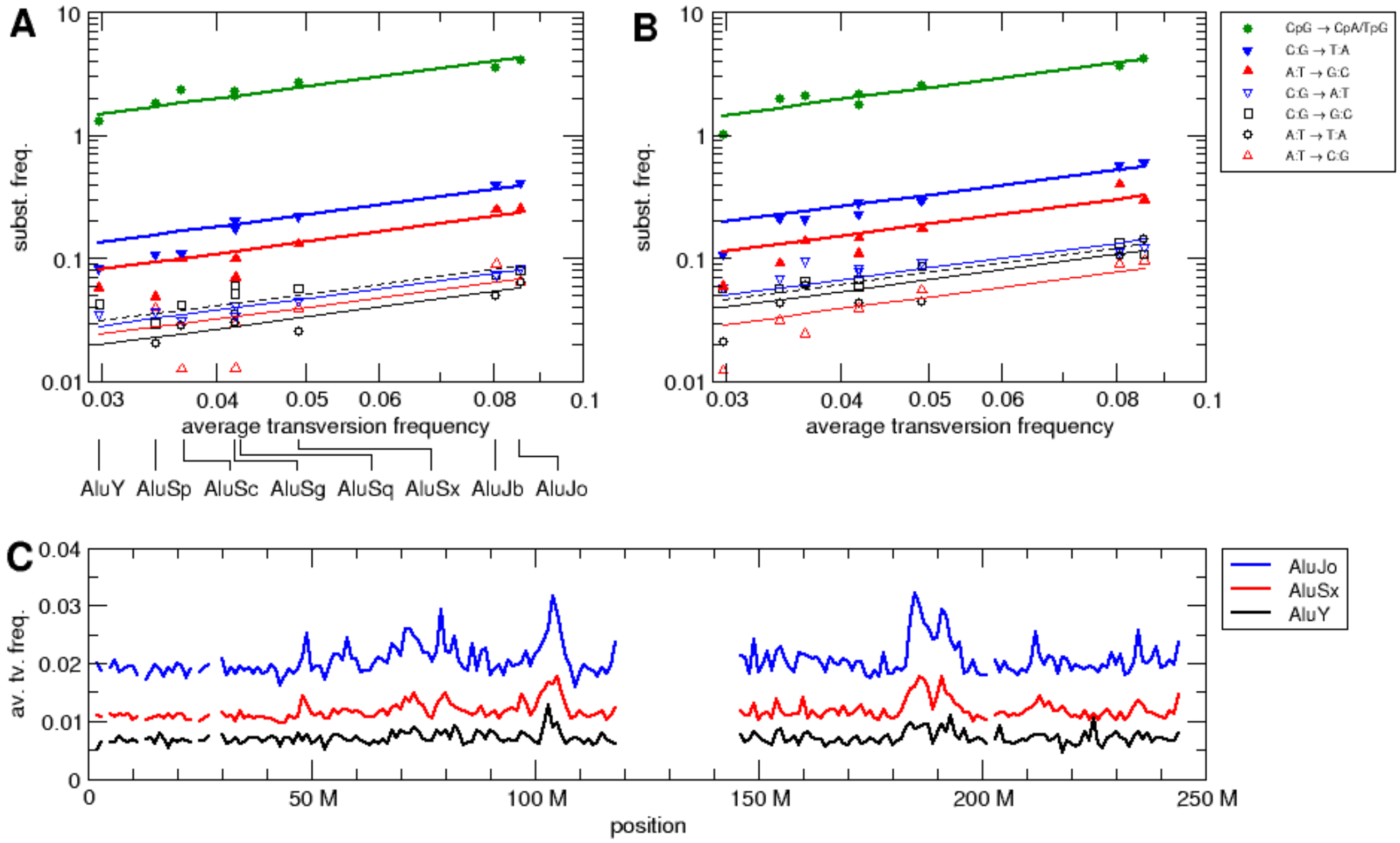

**Figure 1.** The top panels show the transition and transversion frequencies for the 8 Alu families measured in a region of 1Mbp. The substitution frequencies are plotted against the average genome-wide transversion frequency or age on horizontal axis. The two exemplary regions on chromosome 1 are (A) 180-181 Mbp and (B) 191-192 Mbp. The lines represent fitted linear functions ( $y = mx$ ) whose slopes are identified with the relative substitution frequency. A log-log scale has been chosen to represent data, which varies about two orders of magnitude. The fitting was done on a linear scale. (C) The chromosomal profiles of the average transversion frequency for three families of Alus along chromosome 1.

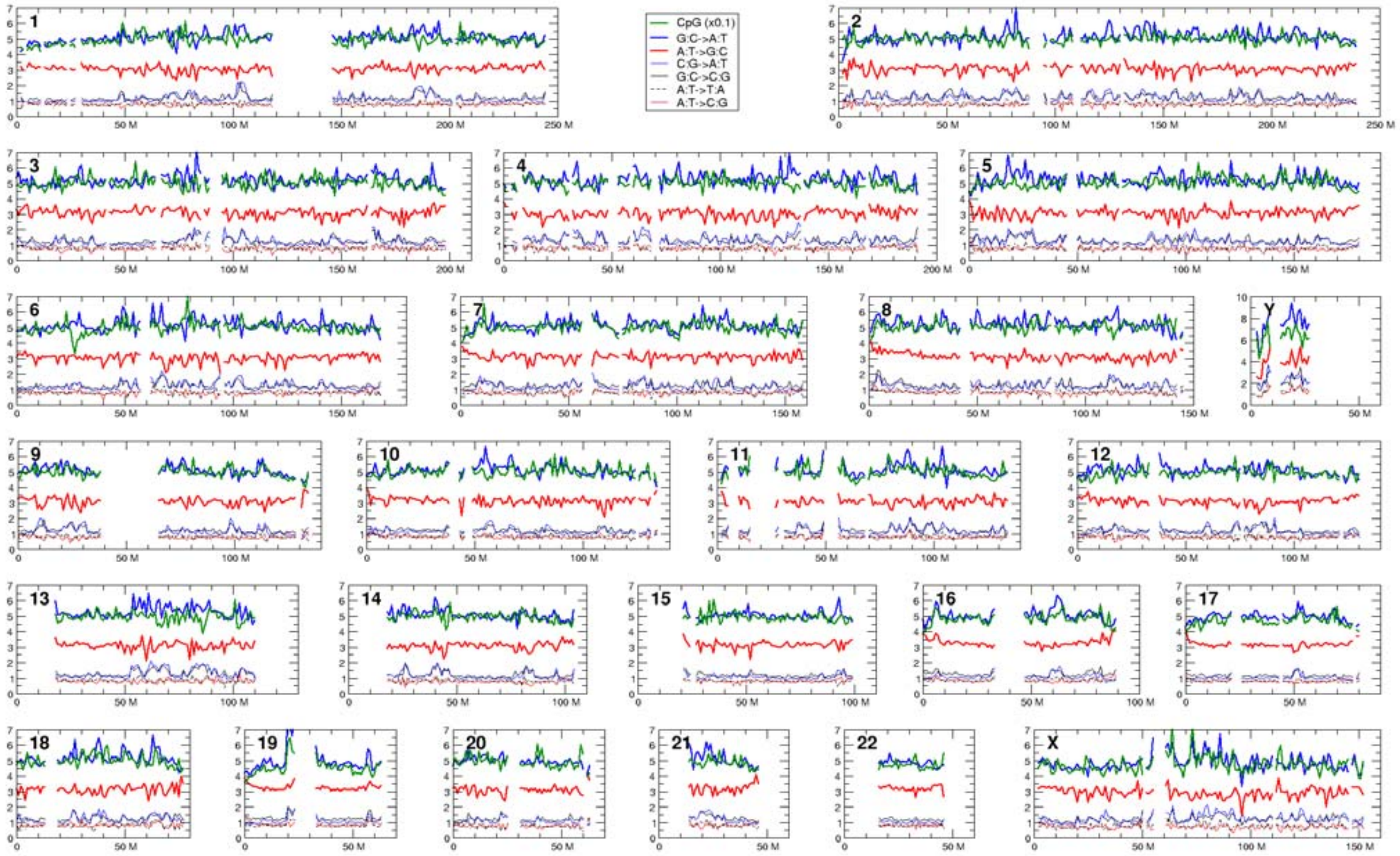

**Figure 2.** Chromosomal profiles of the seven substitution frequencies for all human chromosomes. The frequencies for the CpG-methylation-deamination process have been scaled by 0.1.

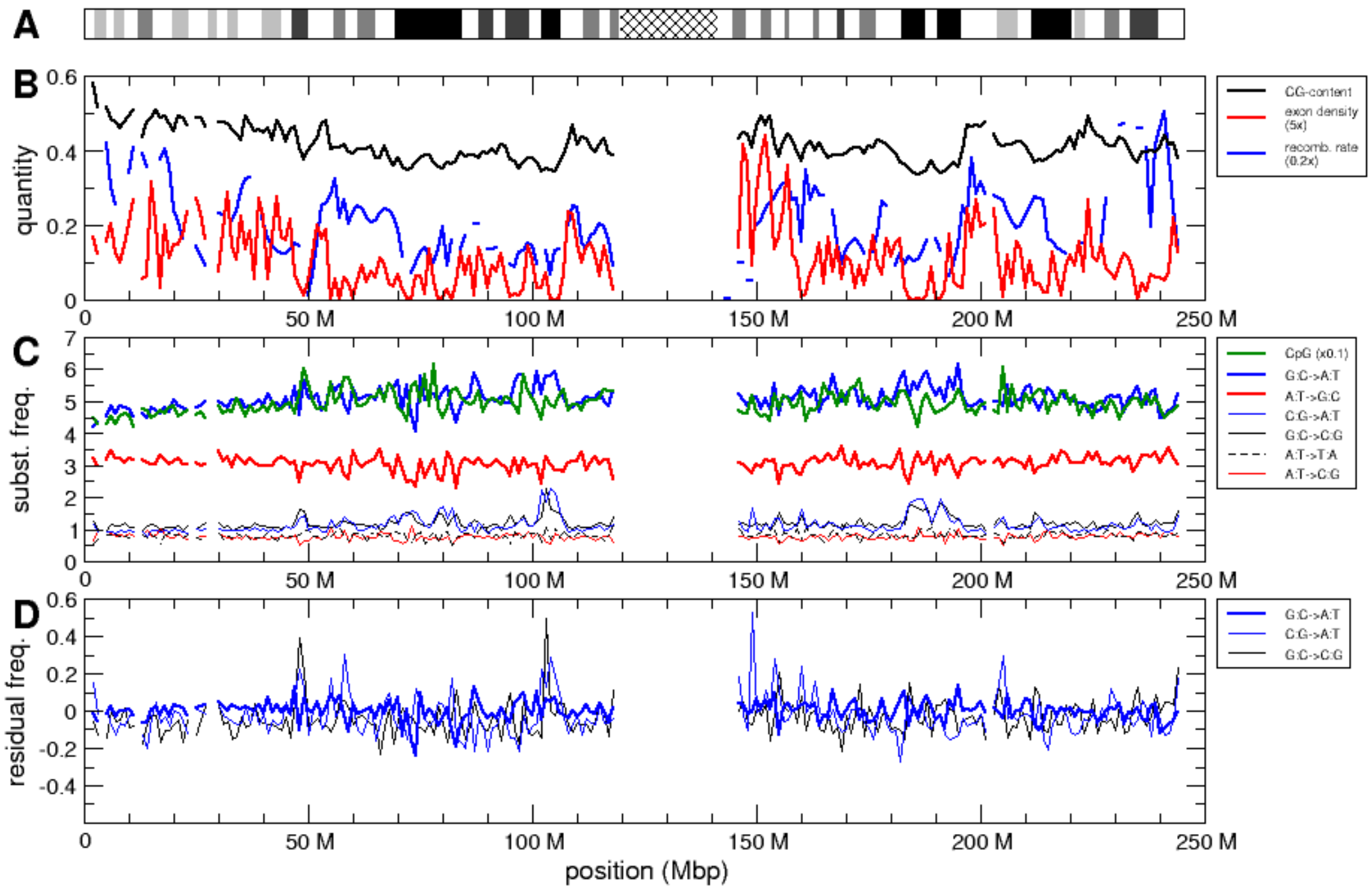

**Figure 3.** Chromosomal profiles of (A) the Giemsa staining density, (B) GC-content, exon density, and recombination rate (C) substitution frequencies, and (D) residual substitution frequencies along the human chromosome 1. Corresponding profiles for all other human chromosomes can be found in the supplementary material.

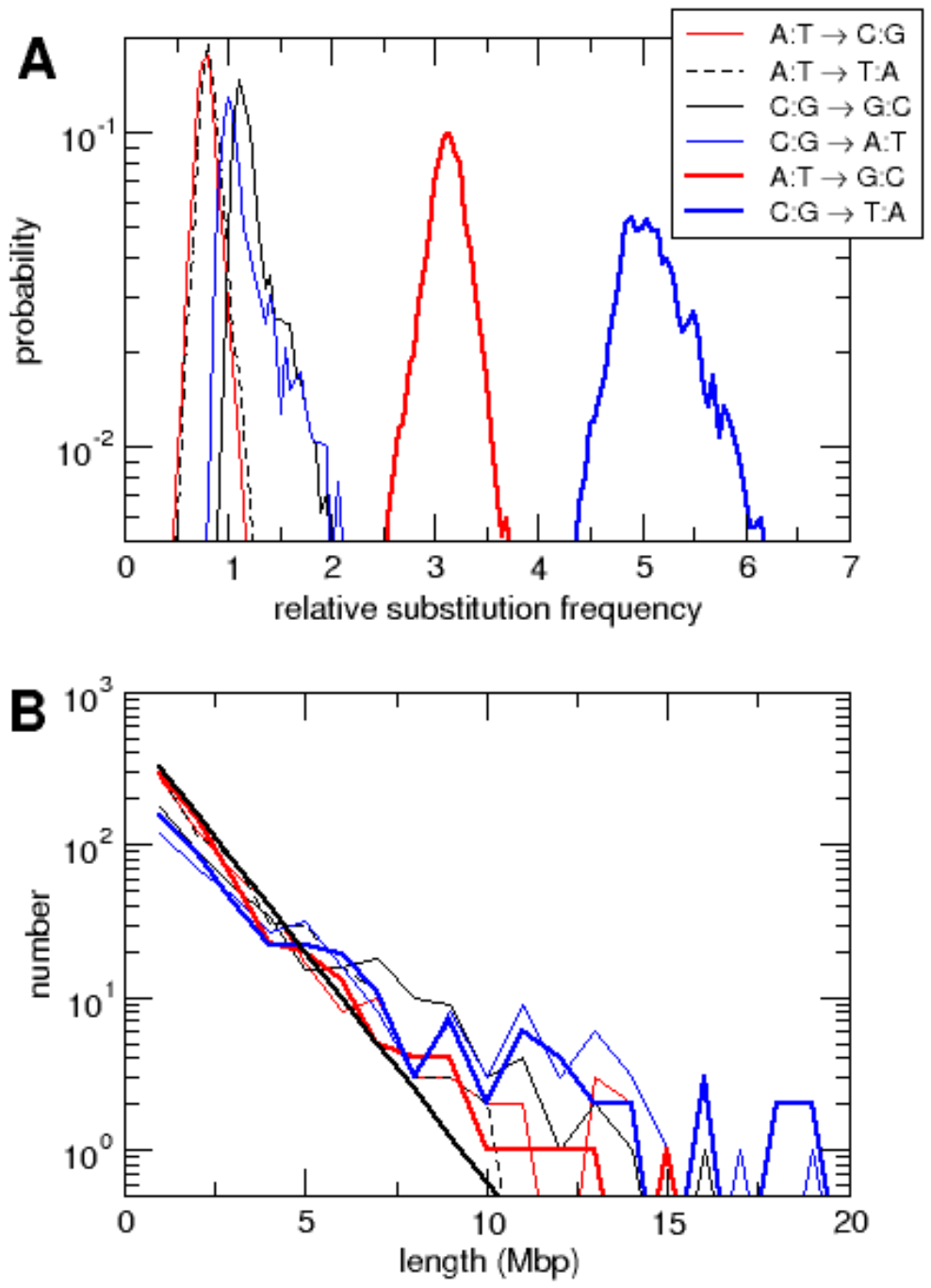

**Figure 4.** (A) Distribution of single nucleotide substitution frequencies in 1 Mbp regions. All shown distributions are skewed and show a significant asymmetric distribution of frequencies with respect to their mode (p<0.0001 for all cases.). (B) The length distributions of regions with substitution frequencies higher than the respective autosomal median. The thick black line represents the expected length distribution of independent regions of length $n$ which is given by $N_0 / 2^{n+2}$ and where $N_0$ is the total number of regions.

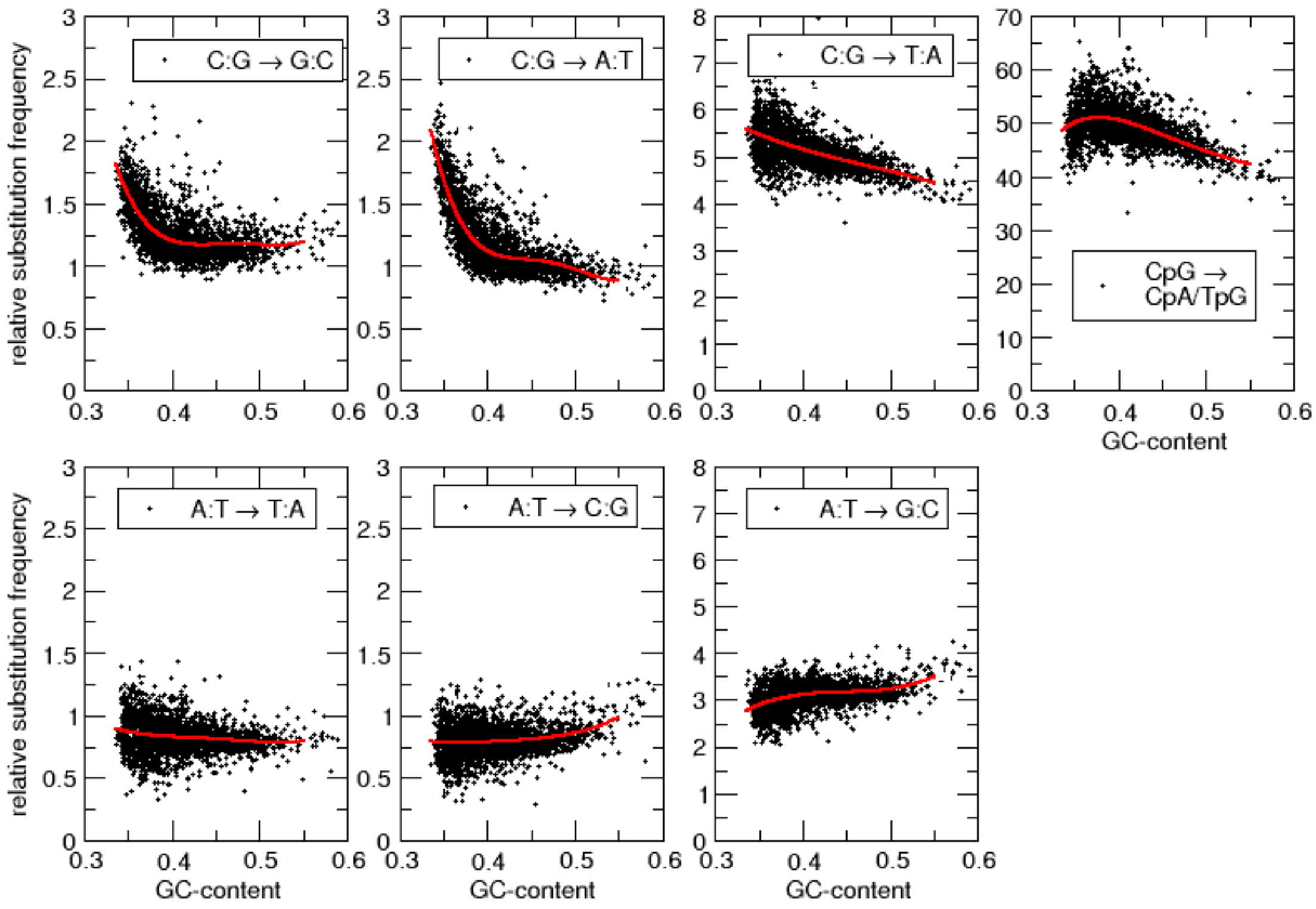

**Figure 5.** Scattergramms of the 7 substitution frequencies with the GC-content. The lines give 4th order polynomial fits to the data in the region from 35–55% GC. The corresponding coefficients can be found in Table 6.

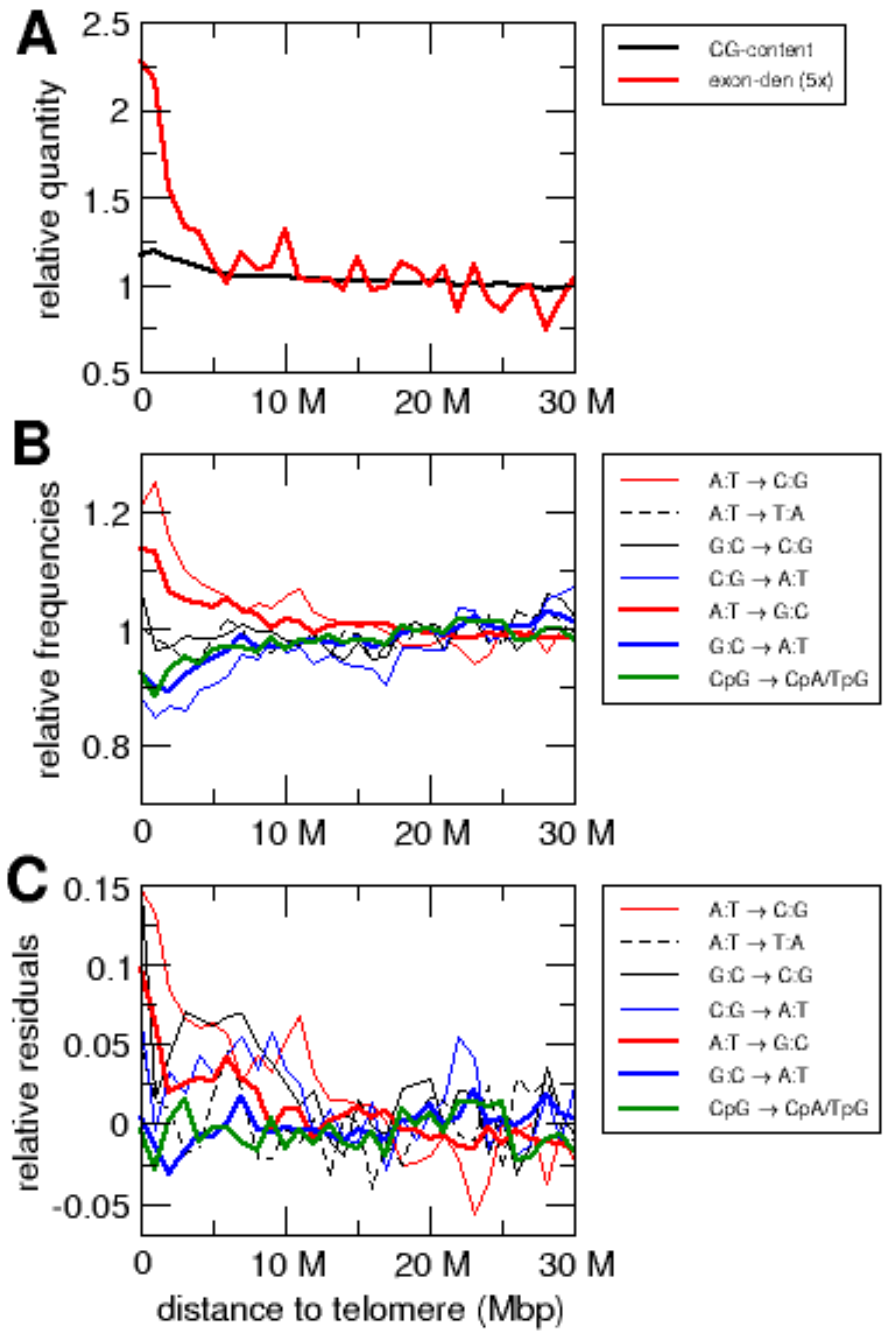

**Figure 6.** Profiles of averaged quantities as a function of the distance to the telomere. (A) GC-content and exon density relative to their genome-wide average, (B) substitution frequencies relative to their genome-wide average and (C) residual substitution frequencies.